\documentclass[prfluids,aps,reprint,superscriptaddress,longbibliography]{revtex4-1}
\usepackage{amsmath,amsfonts,amssymb,amsthm,amsbsy,bm,calrsfs,graphicx,float}
\usepackage{dcolumn}
\usepackage{color}
\usepackage{soul}
\usepackage[colorlinks=true,citecolor=blue]{hyperref}

\newcommand{\bs}[1]{\bm{#1}}
\newcommand{\pa}[1]{\textcolor{cyan}{#1}}

\usepackage{xcolor}

\newcommand{\AAA}{\textrm{\AA}}

\graphicspath{{pics/}}

\begin{document}

\title{Resolving the microscopic hydrodynamics at the moving contact line}
\author{Amal K. Giri}
\author{Paolo Malgaretti}
\affiliation{
Forschungszentrum Jülich GmbH, Helmholtz Institute Erlangen-Nürnberg for Renewable Energy (IEK-11), Cauerstr.\,1, D-91058 Erlangen, Germany}
\author{Dirk Peschka}
\affiliation{Weierstrass Institute Berlin, Mohrenstr.\,39, D-10117 Berlin, Germany}
\author{Marcello Sega}
\email{m.sega@ucl.ac.uk}
\altaffiliation[Present address: ]{Department of Chemical Engineering, University College London, London WC1E 7JE, United Kingdom}
\affiliation{
Forschungszentrum Jülich GmbH, Helmholtz Institute Erlangen-Nürnberg for Renewable Energy (IEK-11), Cauerstr.\,1, D-91058 Erlangen, Germany}
\date{\today}

\begin{abstract}
\noindent

The molecular structure of moving contact lines (MCLs) and the emergence of a corresponding macroscopic dissipation have made the MCL a paradigm of fluid dynamics. Through novel averaging techniques that remove capillary waves smearing we achieve an unprecedented resolution in molecular dynamics (MD) simulations and find that they match with the continuum description obtained by finite element method (FEM) down to molecular scales. This allows us to distinguish dissipation at the liquid-solid interface (Navier-slip) and at the contact line, the latter being negligible for the rather smooth substrate considered.

\end{abstract}

\maketitle

\noindent{}Controlling the properties of MCLs is key to many modern technological applications including printable photovoltaics~\cite{berry_perovskite_2017},  ink-jet printing~\cite{singh_inkjet_2010}, liquid coating and paint drying processes~\cite{deegan_contact_2000}. 
However, modeling the MCL is complicated because continuum descriptions often fail to describe the fluids microstructure in its proximity.
A naive approach based on the no-slip boundary condition encounters  
a non-integrable stress singularity, raising what is known as the Huh-Scriven paradox~\cite{huh_hydrodynamic_1971}. 
Several mechanisms have been proposed to remove the singularity at the MCL 
present in dynamic wetting~\cite{neto_boundary_2005,bonn_wetting_2009,snoeijer_moving_2013,lu_critical_2016}. Among these mechanisms are, for example, the Cox-Voinov law~\cite{voinov_hydrodynamics_1976,cox_dynamics_1986}, the molecular-kinetic theory~\cite{blake_kinetics_1969,blake_influence_2002} and the interface formation theory~\cite{shikhmurzaev_moving_1997,shikhmurzaev_moving_1993}. Notably, all these approaches rely on the introduction of a 
microscopic length scale below which the classical hydrodynamics of a homogeneous, incompressible, Newtonian fluid fails.
To extend the applicability of hydrodynamics, additional dissipation mechanisms have been suggested, including Navier-slip~\cite{lauga_microfluidics_2005} and MCL dissipation~\cite{de_ruijter_droplet_1999,de_gennes_wetting_1985}. In addition, this picture is further complicated by the number of physical mechanisms that can contribute to the dissipation at the liquid/solid interface~\cite{prigogine_statistical_2003,krim_friction_2012}, including phonons~\cite{cieplak_molecular_1994,tomassone_dominance_1997}, electronic excitations
~\cite{tomassone_electronic_1997}, and charge build-up~\cite{burgo_friction_2013}.  

\begin{figure}
\centering
\includegraphics[width=\columnwidth]{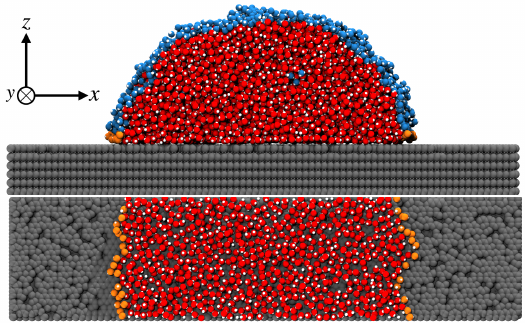}
\caption{MD snapshot of a cylindrical  droplet with 7,332 water molecules in a box 
$L_x\times L_y \times L_z = 198.9\,\AAA\times 46.8\,\AAA \times 170\,\AAA$ and external acceleration of $5\times 10^{-3}\AAA/$ps$^2$ along the $x$-direction. Red: oxygen atoms; Blue: oxygen atoms at the liquid/vapor interface; Orange: oxygen atoms at the MCL; Gray: carbon atoms of the substrate. Top: side view; Bottom: top view of the first water layer\label{fig:snaps}}
\end{figure}

Accordingly, a detailed picture is needed to understand  
these phenomena at the molecular scale in the MCL region. MD simulations can deliver a microscopically detailed picture of the MCL~\cite{qian_molecular_2003} and have been successful on many fronts, for example, in evidencing the breakdown of local hydrodynamics~\cite{thompson_simulations_1989}, providing support for the molecular-kinetic theory both with simple~\cite{bertrand_influence_2009}  and hydrogen-bonding substrates~\cite{lacis_steady_2020} or investigating the time scale for the liquid/solid tension relaxation~\cite{lukyanov_relaxation_2013}. In previous MD studies, the fluid flow at the contact line was affected  by evaporation \cite{freund_atomic_2003} and significant contact line dissipation emerged from strong hydrogen bonding at the liquid-solid interface~\cite{johansson_molecular_2018}, neither of which plays a role in our work.

Despite the potential for providing information at the molecular scale, the presence of thermal capillary waves~\cite{rowlinson_molecular_2002}
is setting a limit to the resolution one can achieve close to the liquid/vapor interface.
Capillary wave theory~\cite{buff_interfacial_1965} predicts that spontaneous surface excitations broaden the width of fluid interfaces with a logarithmic divergence that depends on the simulation cell size. For small molecular liquids and typical simulation cells, these fluctuations are larger than the molecular size~\cite{rowlinson_molecular_2002,chacon_intrinsic_2003} at room temperature. Several computational strategies have been devised to recover the intrinsic structure of fluid interfaces~\cite{werner_anomalous_1997,werner_intrinsic_1999,chacon_intrinsic_2003, jorge_intrinsic_2007,partay_new_2008, willard_instantaneous_2010,sega_generalized_2013}. In some of these, including the approach reported here, one first identifies molecules at the phase interface as shown in Fig.~\ref{fig:snaps}, then uses them to define a local coordinate system. In turn, one can use this coordinate system to compute the profiles of chosen observables as a function of the local distance from the interface. 
To obtain further insight into the dynamics of the molecules at the MCL, one has to cope with its fluctuations. In this case, however, two interfaces are involved (liquid/vapor and liquid/solid), and one-dimensional intrinsic profiles are not enough to describe the system. 
Here, we extend the approach of intrinsic profiles to the region around the three-phase contact line by introducing two-dimensional density maps that are functions of the position relative to the liquid/vapor and liquid/solid interfaces, respectively.
This procedure allows us to resolve the structure and the flow close to the MCL at an unprecedented resolution.

We performed MD simulations of a cylindrical water droplet moving on a rigid substrate under the influence of a constant acceleration parallel to the solid surface.
%
Water is modeled using the SPC/E potential\cite{berendsen_missing_1987} and the substrate as a rigid, graphite-like structure with defects~\cite{sega_regularization_2013}, obtained by removing 35\% of randomly selected surface atoms.
The presence of defects serves rather well the purpose of providing a  friction coefficient thanks to the lateral modulation of the water-surface interaction potential. All molecular dynamics simulations are performed with an in-house modified version of the GROMACS simulation package, release 2019.4\cite{abraham_gromacs_2015}, that  uses a Nos\'e-Hoover \cite{nose_molecular_1984,hoover_canonical_1985} thermostat coupled only to the direction orthogonal to both the surface normal and the external force\cite{thompson_simulations_1989}, keeping in mind that this introduces a bias in the surface energies \cite{sega_impact_2018}. Further details are reported in the Supplemental Material\cite{supplementary}\nocite{abraham_gromacs_2015,berendsen_missing_1987, sega_regularization_2013, nose_molecular_1984,hoover_canonical_1985, thompson_simulations_1989,sega_impact_2018, miyamoto_settle_1992,essmann_smooth_1995,michaudagrawal_mdanalysis_2011,sega_pytim_2018,sega_generalized_2013,sega_pytim_2018,shvab_intermolecular_2013,sega_long-range_2017,sega_long-range_2017}. Input files and GROMACS source code patch are provided in a dataset available on Zenodo\cite{dataset}.

The first step in computing intrinsic maps of the droplet's hydrodynamic fields is identifying the surface molecules at the liquid/vapor interface,  cf.~Fig.~\ref{fig:snaps}. 
Here, as detailed in the Supplemental Material\cite{supplementary}, we use the GITIM algorithm~\cite{sega_generalized_2013} as implemented in the Pytim package\cite{sega_pytim_2018}.
\begin{figure}
    \includegraphics[trim=0 0 0 0, clip,width=\columnwidth]{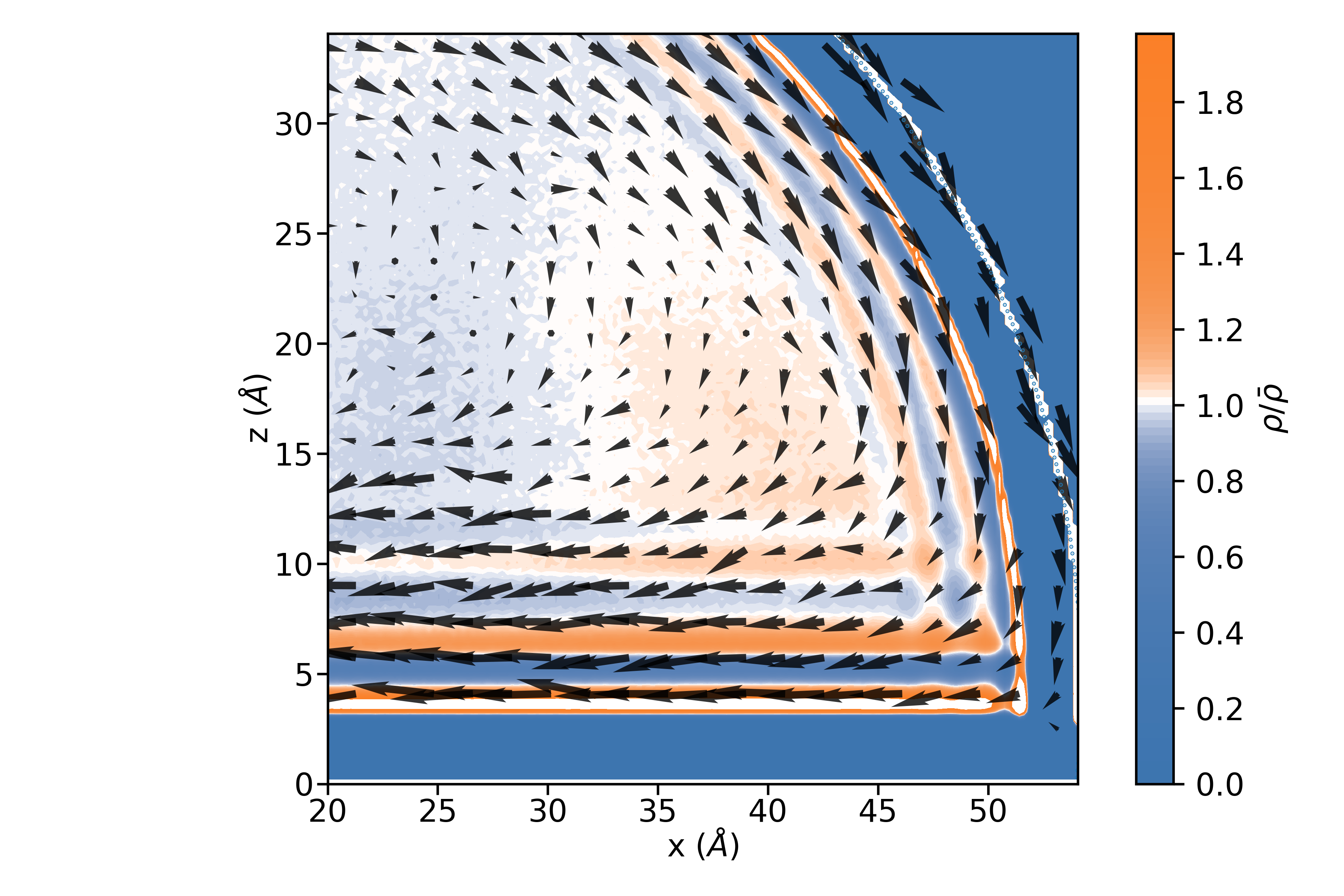}
    \caption{Intrinsic map of the density (color) and velocity field in the co-moving reference frame (arrows). The molecules belonging to the liquid/vapor interface have zero inherent distance \(\xi_V\) and are represented here using the result of an elliptic fit through their average position (small open circles).
    The density field has resolution  \(0.22\,\AAA\times{}0.2\,\AAA\) and is normalized with respect to the bulk density. The velocity field has resolution  \(1.7\,\AAA\times{}1.6\,\AAA\)  and is scaled such that an arrow of length 1 \(\AAA\) on the plot corresponds to a velocity of 0.01 \(\AAA\)/ps. \label{fig:flow}}
\end{figure}
The surface molecules are used to provide the location of the (continuum) liquid/vapor \(\bm{\zeta}_V(\bm{r})\) and liquid/solid \(\bm{\zeta}_S(\bm{r})\) interfaces by using a linear interpolation between neighboring triplets of surface molecules.  We compute the intrinsic map of a generic local observable \(A\) as 
\( A(\bm{\xi})  =
    \left\langle \sum_i A_i\prod_\alpha \delta\left(\xi_\alpha-\min\left|\bm{r}_i-\bm{\zeta}_\alpha\right|\right) \right\rangle/ {N(\bm{\xi})}.
\)
Here, \(\alpha=\{V,S\}\) labels the liquid/vapor and liquid/solid interfaces and \( \bm{\xi} = ( \xi_V,\xi_S)\) are the distances of a point in space from the two surfaces. The angular brackets indicate a statistical  average over time, and the index \(i\) labels particles.
The normalization factor \(N(\bm{\xi})\) is calculated from the intrinsic density map of uniformly distributed random points within the liquid phase.
The two-dimensional intrinsic map \(A(\bm{\xi})\)
is best interpreted if remapped back from the generalized coordinates \((\xi_V,\xi_S)\) to the Cartesian ones using the average location of the interfaces as the origin of the local frame.  This approach provides precise information close to the liquid/vapor interface but loses accuracy far away from the interface.
However, this region is not of concern for the current investigation, which focuses on the neighborhood of the MCL.

\begin{figure}[t]
    \centering
    
    \includegraphics[ width=\columnwidth]{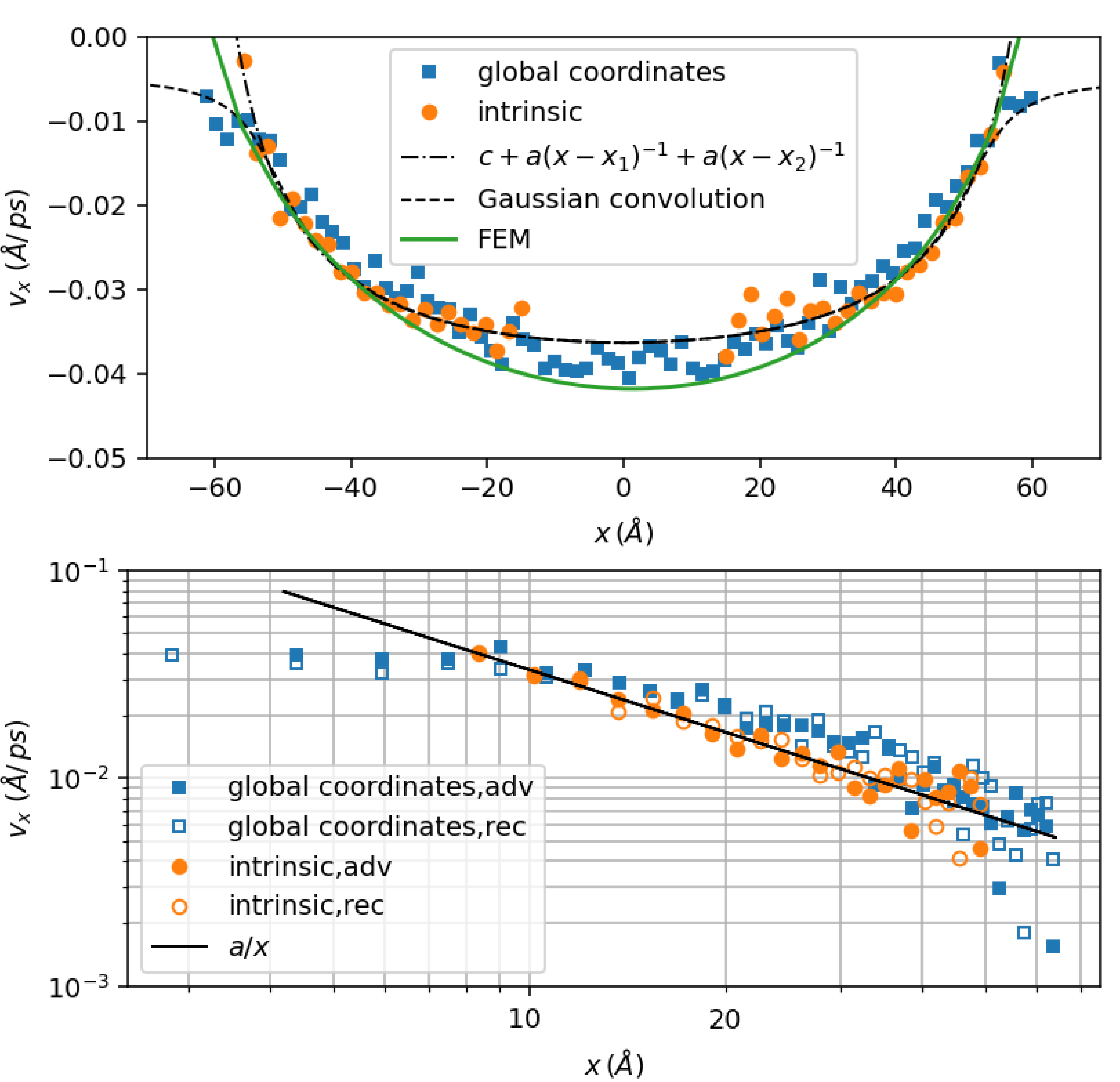}  
    \caption{Top panel: horizontal velocity profiles (squares: global coordinate system, circles: intrinsic profiles) sampled along \(x\) at \(z = 3\pm3\,\AAA\). The dot-dashed curve is a fit to a power law \(c+ a(x-x_1)^{-1} + a
    (x_2-x)^{-1}\). 
    The dashed line is the convolution of the best-fit power law with a Gaussian as described in the text (\(a=-0.435\,\AAA{}^2/\mathrm{ps}\), \(c=-0.0495\,\AAA/\mathrm{ps}\), \(x_1=66.42\,\AAA\), \(x_2 = -66.33\,\AAA\)). 
    The green curve is the FEM solution evaluated at the interface \(z=0\). 
    Bottom panel: double logarithmic scale, same data as in the top panel but symmetrized along the vertical axis and shifted so that the fitting constants \(c\) and \(x_1\) are zero (\(a=0.335\,\AAA{}^2/\mathrm{ps}\)).} \label{fig:cuts}
\end{figure}

\begin{figure*}
\centering
\includegraphics[width=\textwidth]{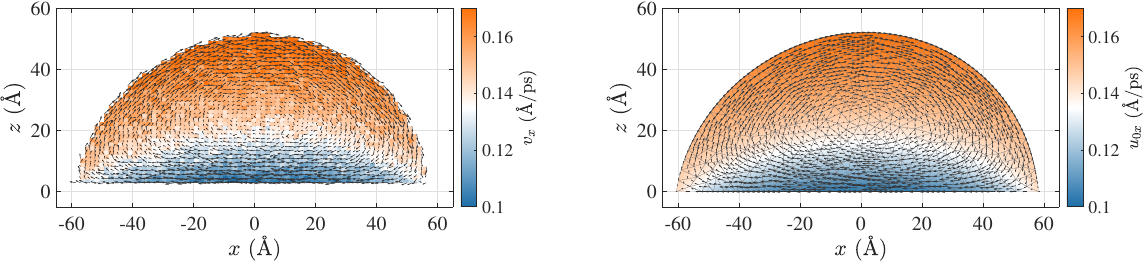}
\caption{(Left: averaged MD velocity profile $\bs{v}=(v_x,v_z)$ (shown for $\rho>0.66\bar{\rho}$); Right: optimal Stokes velocity profile $\bs{u}_0=(u_{0x},u_{0z})$ for $\beta=1.38$ (slip-length $b= 40\AAA)$ and $\delta=0$, $\mathrm{Bo}=0.2414$ and $\vartheta_\mathrm{e}=97.2^\circ$ in a comoving frame (vectors) and absolute horizontal velocity (shading).} 
\label{fig:flowMDvsST}
\end{figure*}

The intrinsic density field reported in Fig.~\ref{fig:flow} reveals the oscillations associated to molecular layering at both the liquid/solid and liquid/vapor interfaces. 
In particular, a pronounced gap region due to hard-core repulsion appears next to both interfaces.
As shown in Fig.~\ref{fig:flow}, the resolution of the velocity field matches the molecular scale, with bins of size \(1.7\,\AAA \times 1.6\,\AAA\), while the density bins are of size  \(0.22\,\AAA\times0.2\,\AAA\). 
We do not observe any trace of an influence of positional correlations on the velocity field, which is a much smoother function of the position than the density. It is often remarked that continuum hydrodynamics breaks down in simple liquids below scales of few molecular diameters \cite{chung_generalized_1969,ailawadi_generalized_1971}.
However, this is true for processes like sound propagation that, at that scale, happen during extremely short times on the order of a picosecond.
In contrast, we compute the stationary fields of the hydrodynamic quantities by averaging over time intervals so large that no memory effect can survive.

The density map is in fact a correlation function and the oscillations can be interpreted as in a pair correlation function. The local density, computed as the (reciprocal) average volume available for each atom, is constant over the whole liquid phase and the droplet is incompressible (see plot of $\nabla\cdot\mathbf{v}$ in Fig. A8 of the Supplemental Material\cite{supplementary}) down to and below the molecular scale. Therefore, the stationary velocity field that we compute from MD data can be modeled as that of a homogeneous incompressible fluid in the FEM simulations.

In Fig.~\ref{fig:cuts} we report the horizontal velocity profile for molecules in the slab \(z<6\AAA\), computed in the droplet co-moving frame using both the global and intrinsic coordinate systems. 
The velocity sampled in the global coordinate system extends further than the intrinsic one because of the fluctuations. 
Both profiles coincide in the region \(|x|<50\AAA\), showing a steep increase when departing from the droplet center.
However, close to the MCL, the velocity profile computed in the global coordinate system flattens out. In contrast, the intrinsic one is compatible over its entire extension with a 
power-law slip profile.
The profile in the global coordinate system can be very well described in the spirit of the capillary wave theory by using quantities \(\widetilde{A}(x) = \int_{0}^{L_x} G(x-x') A(x') {\rm d}x'\) convoluted  with the  Gaussian distribution \(G(x)\sim{}\exp[-x^2/(2\sigma^2)]\) of the interface fluctuations, so that \(v_x = \widetilde{p}_x/\widetilde\rho\). Here,  $A(x)$ is the profile of observable $A$ averaged over the  whole range in $y$ direction and over  the first molecular layer in $z$ direction.

In the top panel of Fig.\,\ref{fig:cuts} we report as a dashed-line the velocity profile obtained by applying this convolution procedure to the best fit intrinsic velocity profile 
\(v_x\sim (x-x_1)^{-1} + (x_2-x)^{-1}\) and a step-like density profile, using the interfacial layer width \(\sigma = 2.5\AAA\).
The resulting profile reproduces very well the flattening in the region close to the MCL. 
These features are more evident in the double logarithmic scale, reported in the right panel of Fig.\,\ref{fig:cuts}, where we have symmetrized the profile around its minimum and shifted it vertically and horizontally by the fitting constants \(x_1\)  and \(c\).
Our results show that after removing the capillary fluctuations via the intrinsic map method, the power-law behavior extends over the entire range down to the molecular length scale, and the previously reported departure from the power-law\cite{qian_power-law_2004} is due to the 
smearing
introduced by capillary waves rather than a breakdown of hydrodynamics at short length scales.

The present computational approach allows us to discuss some of the proposed corrections to continuum models. We compare our MD simulation results with a full hydrodynamic continuum model for the MCL, where we solve a Stokes problem for an incompressible viscous Newtonian fluid (Re=0) featuring a free interface subject to capillary forces, which is driven by a constant acceleration $\bs{g}$. Dissipative processes are included by the combination of Navier-slip, i.e., $\beta u_x = \mu \partial_z u_x$, and dynamic contact line dissipation \cite{ren_boundary_2007}, i.e.,  $\nu_x\cdot u_x = \gamma/\delta(\cos\theta_\mathrm{e}-\cos\theta)$. 
The slip parameter $\beta$  enhances the inhomogeneity in the flow by increasing $\partial_z u_x$. In contrast, the contact line friction $\delta$ enhances the asymmetry of the droplet shape through advancing and receding contact angle, e.g., cf. Fig.~A5 of the Supplemental Material\cite{supplementary}.
Note that we use constant slip length and contact line dissipation, as space-dependent coefficients would yield an overdetermined improved fit but not lead to new physical insights.
We discretize the system of partial differential equations using the isoparametric $\bs{P}_2-P_1$ Taylor-Hood finite elements combined with an ALE mesh motion strategy, for details cf. the Supplemental Material\cite{supplementary} and Refs. \cite{peschka2018variational,montefuscolo2014high}.

\begin{figure}[t]
\centering
\includegraphics[width=0.8\columnwidth]{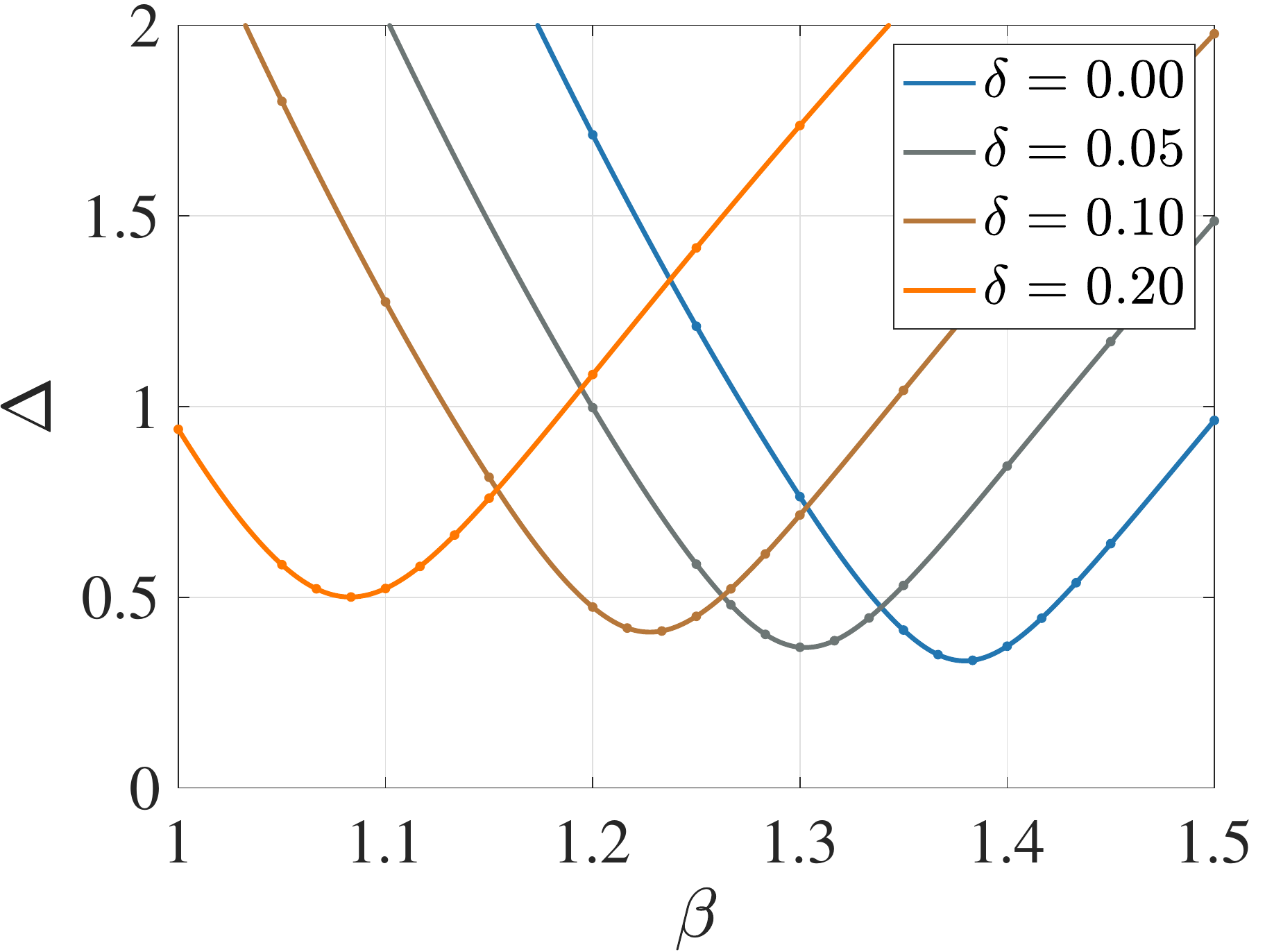}
\caption{Rmsd $\Delta$ between FEM and MD flow fields for different  dissipation parameters $\beta$ (Navier-slip) and $\delta$ (contact line) in the hydrodynamic Stokes model.}
\label{fig:RMSD}
\end{figure}

For a given acceleration $\bs{g}$ and droplet size $L$, we compare the velocity of traveling-wave solutions from the Stokes problem 
$\bs{u}(t,\bs{r})=\bs{u}_0(x-v_0t,z)$ with the corresponding velocity from the MD problem by computing the root mean square deviation (rmsd) \(\Delta\), relative to the MD accuracy, with 
\(
 \Delta{}^2 = \int_{\Omega_0} \rho|\bs{u}_0-\bs{v}|^2/\delta\bs{v}^2{\,\rm d}x{\,\rm d}z  \big/ \int_\Omega \rho{\,\rm d}x{\,\rm d}z
\),
in the comoving frame $\Omega_0=\Omega-(v_0,0)t$, where $\delta\bs{v}(x,z)$ is the space-dependent MD averaging error of $\bs{v}$. In Fig.~\ref{fig:flowMDvsST} we show the excellent agreement of the horizontal components of MD and Stokes velocity for a droplet with optimized $\beta$ and $\delta$. Correspondingly, in Fig.~\ref{fig:RMSD} we show the rmsd $\Delta$ for different pairs of $\delta$ and $\beta$, where the optimal parameters suggest a vanishing or small contact line dissipation $0<\delta<0.05$ and $\beta=1.38$, corresponding to a slip-length $b=40\,\AAA$. 
Because $\Delta$ is a relative measure of inaccuracy with respect to that of MD, we expect $\Delta\sim \mathcal{O}(1)$ for a good fit, similarly to a reduced $\chi^2$ test. For given $\delta$, the local minima in Fig.~\ref{fig:RMSD} suggest that also other parameters with larger $\delta$ and smaller $\beta$ are feasible. However, those parameters are not optimal for the prescribed bulk viscosity $\mu=0.7\,{\rm mPa\,s}$ but could possibly be improved  by fitting the entire triple $(\mu,\beta,\delta)$, considering spatially  inhomogeneous parameters or a compressibility of the fluid.

\begin{figure}[t]
    \centering
    \includegraphics[trim=10 10 0 10, clip,width=\columnwidth]{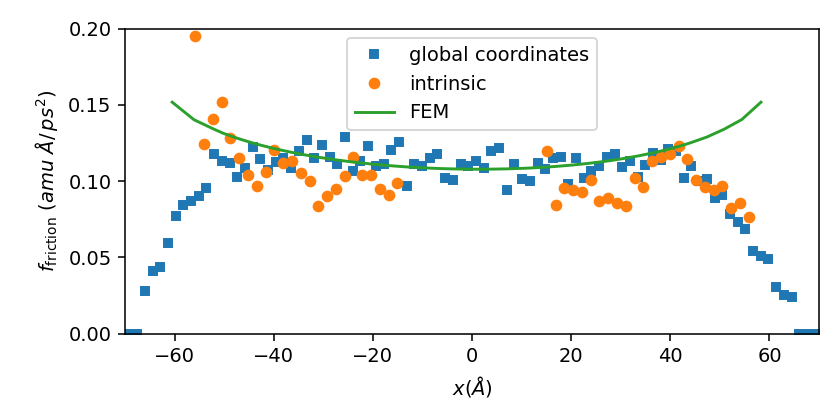}
    \caption{Profile of the friction surface density computed from the MD data in the global (squares) and intrinsic (circles) coordinate systems.
     The green curve is the FEM solution evaluated at the interface \(z=0\). } 
    \label{fig:friction}
\end{figure}

The water molecules are subjected to a friction force that, in steady-state conditions,  balances \(\bs{g}\) and is defined as the horizontal component of the force density  \(f_{\Gamma, x}(x,z)\) acting from the substrate on water molecules. 
Even though long-range corrections for dispersion forces are in place, the friction force becomes negligible at larger elevations 
\(z\gtrapprox10\,\AAA\)
and is acting almost exclusively on the first layer of water molecules at the substrate. Therefore, we can define a friction force per unit area as $F_\Gamma(x)  = \int_0^\infty f_{\Gamma, x}(x,z) {\rm d}z$.
This force is the counterpart of the surface friction that in sharp interface models like the present FEM can be calculated from the stress $\bs{t}\cdot\bs{\sigma}\bs{\nu}\equiv -\mu_\Gamma\bs{t}\cdot\bs{u}$ at the solid-liquid interface, where $\bs{t}$ and $\bs{\nu}$ are the corresponding tangential and normal vectors.
Fig.~\ref{fig:friction} shows the friction surface density \(F_\Gamma\) obtained from the MD trajectories and compared to the FEM results. Far from the droplet edges, both force profiles are in excellent agreement. However, close to the advancing and receding MCL, the friction force behaves qualitatively differently. This mismatch contrasts the overall good match of the slip velocity profile Fig.~\ref{fig:cuts}. Therefore, even though the Navier-slip boundary condition is an excellent effective model in reproducing the slip velocity, the same is not true concerning the friction force near the MCL. 

This difference implies that the underlying dynamics is more complex than that encoded into the Navier-slip boundary condition. Remarkably, even in the present case, where the substrate is particularly smooth and the optimal FEM solution is compatible with zero contact line dissipation, we observe a significant deviation of the friction force from the one emerging from the Navier-slip boundary condition. Note that, upon integration of the force density in Fig.~\ref{fig:friction}, this contribution is compatible with a contact line friction $\delta \lessapprox 0.05$. Larger defects or soft substrates will enhance this effect.

In this work, we compared microscopic MD simulations and hydrodynamic FEM simulations of sliding droplets with respect to their densities, flow and force fields at interfaces and MCLs. Using a novel averaging method, we showed that assumptions of continuum hydrodynamics are valid near the MCL. However, despite an effectively vanishing contact line friction, corrections to the force balance are clearly visible. In the future, larger droplets or elastic substrates with stronger roughness should be investigated to further enhance contact line friction and suppress density oscillations. Additionally, one should consider thermodynamics, compressibility, and evaporation of the liquid phase to better match continuum hydrodynamics with the microscopic interacting-particle system.

The datasets available on Zenodo\cite{dataset} provide MD input files, the GROMACS source code patch for the thermostat, averaged data from MD simulations (Cartesian mesh with mass density and velocity field) as well as FEM simulation solutions (triangular 2D droplet mesh with velocity field) in ASCII format, and MATLAB functions for the data import and plot generation. The MATLAB code is available at \url{https://github.com/dpeschka/stokes-free-boundary/}.

\section*{acknowledgments}
DP, AKG and MS acknowledge the funding by the German Research Foundation (DFG) through the projects \#422792530 (DP) and \#422794127 (AKG and MS) within the DFG Priority Program SPP 2171 \emph{Dynamic Wetting of Flexible, Adaptive, and Switchable Substrates}.


%
\end{document}